\documentclass[pdflatex,sn-mathphys-num]{sn-jnl}

\usepackage{graphicx}%
\usepackage{multirow}%
\usepackage{amsmath,amssymb,amsfonts}%
\usepackage{amsthm}%
\usepackage{mathrsfs}%
\usepackage[title]{appendix}%
\usepackage{xcolor}%
\usepackage{textcomp}%
\usepackage{manyfoot}%
\usepackage{booktabs}%
\usepackage{algorithm}%
\usepackage{algorithmicx}%
\usepackage{algpseudocode}%
\usepackage{listings}%
\usepackage{bm}

\newcommand{\TITLE}{The role of the effective range in strongly-interacting few-body systems}

\begin{document}

\title[\TITLE]{\TITLE}

\author*[1]{\fnm{Lucas} \sur{Madeira}}\email{madeira@ifsc.usp.br}

\affil*[1]{\orgdiv{Instituto de Física de São Carlos}, \orgname{Universidade de São Paulo}, \orgaddress{\street{Av. Trabalhador Sancarlense, 400}, \city{São Carlos}, \postcode{CP 369, 13560-970}, \state{São Paulo}, \country{Brazil}}}

\abstract{
Strongly interacting systems appear in several areas of physics and are characterized by attractive interactions that can almost, or just barely, loosely bind two particles. Although this definition is made at the two-body level, this gives rise to fascinating effects in larger systems, including the so-called Efimov physics.
In this context, the zero-range theory aims to describe low-energy properties based only on the scattering length. However, for a broad range of physical applications, the finite range of the interactions plays an important role. In this work, I discuss some aspects of finite-range effects in strongly interacting systems. I present the zero-range and shapeless universalities in two-body systems with applications in atomic and nuclear physics. I derived an analytical expression for the $s$-wave bound-state spectrum of the modified P\"oschl-Teller potential for two particles in three dimensions, which is compared with the approximations to illustrate their usefulness. Concerning three identical bosons, I presented a trimer energy scaling function that explicitly includes the effective range. The implications for larger systems are briefly discussed.
}


\maketitle

\section{Introduction}
\label{sec:intro}

In quantum systems near unitarity, i.e. diverging two-body scattering length, the particles are distributed in spatial scales larger than the interaction range, making the specific interparticle potential less crucial for reproducing the ground-state spectrum~\cite{Braaten2006,Braaten2007,Naidon2017}. This understanding is key to explaining phenomena like the Thomas collapse~\cite{Thomas1935} and the Efimov effect~\cite{Efimov1970,Efimov1981}, where the former involves the collapse of the three-body ground state as the interaction range diminishes, and the latter indicates an infinite number of three-body bound states at the unitary limit. Both phenomena are linked through a scale transformation~\cite{Adhikari1988}.

Bound and resonant states appear as we approach the unitary limit, showing independence from the specifics of the two-body potential~\cite{Newton2013}. This phenomenon, observed by Phillips~\cite{Phillips1968} in his study of the correlation between the triton binding energy and the nucleon-deuteron scattering length, was further explained by Efimov and Tkachenko through the zero-range theory~\cite{EfimovTkachenko1985}.

The study of few-nucleon correlations, which elucidated phenomena like the Thomas collapse and the Efimov effect, is heavily in debt of the pioneering mathematical work on three-body problems by Skorniakov and Ter-Martirosian~\cite{Skorniakov1957}, Danilov~\cite{Danilov1961}, and Faddeev~\cite{Faddeev1965}. Initially linked to renormalization groups~\cite{Albeverio1981}, the Efimov effect is characterized by a unique scaling symmetry~\cite{Amorim1997,Frederico1999} and related to renormalization group limit cycles~\cite{Bedaque1999,Bedaque1999b}. Its significance across both nuclear and atomic few-body systems led the community to expand this universality to more complex systems through experimental and theoretical studies~\cite{Braaten2006,Braaten2007,Frederico2011,Frederico2012a,Hammer2013,Zinner2013,Naidon2017,Greene2017,Kievsky2021} since it highlights the universal behavior in three-body systems with infinite scattering length.

The exploration of Efimov states in physical systems, initially theoretical, expanded into empirical research on configurations of a few nucleons and atoms~\cite{Fonseca1979,Lim1980,Adhikari1982,Adhikari1982b}. Despite challenges in nuclear physics due to nucleon-nucleon interaction properties, identifying exotic nuclear systems with two-neutron halos~\cite{Hansen1995b,Hansen1995} has provided a promising avenue for investigating potential Efimov states~\cite{Fedorov1994,Amorim1997}.

The first prediction of Efimov states in few-atomic systems, specifically in helium gases at low temperatures, was made by Lim et al. in 1977~\cite{Lim1977}. This was followed by further substantiation regarding an excited Efimov state in the three-helium atomic system by Cornelius in 1986~\cite{Cornelius1986}. In 2015, Kunitski et al. experimentally confirmed the excited Efimov state in the $^4$He trimer~\cite{Kunitski2015}. Discussions on the ultracold collision properties of $^4$He trimers have been ongoing~\cite{Kolganova1997}, with recent studies exploring collisions involving a $^4$He dimer and a third atomic particle, including $^4$He, $^{6,7}$Li, and $^{23}$Na, in the context of Efimov physics~\cite{Suno2017,Shalchi2020}.

The 1995 experimental achievement of Bose-Einstein condensation in ultra-dilute atom clouds~\cite{Anderson1995,Davis1995,Bradley1995} marked a significant advancement in atomic physics, further enhanced by the ability to manipulate atom-atom interactions through Feshbach resonances~\cite{Feshbach1958,Feshbach1962,Timmermans1999,Chin2010}. This paved the way for the experimental discovery of Efimov states in atomic systems, with the first evidence found in ultracold cesium atoms~\cite{Kraemer2006}. The ongoing exploration of Efimov states is a focus of both theoretical~\cite{Fermi2023} and experimental research~\cite{Bougas2023,Etrych2023} in the field.

The Efimov effect extends to systems with more than three particles. Notably, Tjon discovered a correlation, known as the Tjon line~\cite{Tjon1975,Tomio2013}, between the binding energies of tetramers and trimers in $^4$He, which connects back to Efimov physics~\cite{Platter2004,Platter2005,Yamashita2006,Ferlaino2009}. Additionally, Coester et al. explored how variations in nuclear-matter binding energy relate to two-body potentials with equivalent phase shifts~\cite{Coester1970}.

The above discussion is focused on the scattering length, but for a broad range of physical systems, the range of the interactions cannot be neglected. In this work, I aim to discuss some aspects of finite-range corrections to properties of strongly interacting few-body systems. The main goal driving the different approaches is to extend the universality region by taking into account the interaction range, the same reasoning behind going from a zero-range theory to a finite-range one in two-body systems.

This work is organized as follows. Section~\ref{sec:two_body} deals with two-body systems. In Sec.~\ref{sec:zero_shapeless_universalities}, the effective range expansion is introduced to motivate the concepts of zero-range and shapeless universalities. Section~\ref{sec:dimers_physical} discusses their applicability to physical systems. In Sec.~\ref{sec:illustration_two_body}, a microscopic two-body potential is used to illustrate both approximations, and it is compared to an analytical expression for the bound-state spectrum of the modified P\"oschl-Teller potential for two particles in three dimensions, derived in Appendix~\ref{sec:app_mpt}. Section~\ref{sec:three_body} deals with three identical bosons, focusing on the formalism introduced in Ref.~\cite{Madeira2021}. Finally, the conclusions are presented in Sec.~\ref{sec:conclusion}, briefly discussing implications for larger systems.

\section{The two-body sector}
\label{sec:two_body}

\subsection{Zero-range and shapeless universalities}
\label{sec:zero_shapeless_universalities}

Low-energy scattering theory allows a universal description of two-body scattering for local finite-ranged spherically symmetric potentials~\cite{Lima2023}. A seminal work by Hans Bethe~\cite{Bethe1949} related the $s$-wave scattering length $a$, the effective range $r_0$, and the $s$-wave phase shift $\delta_0(k)$ through
\begin{equation}
\label{eq:delta}
k \cot \delta_0(k) = -\frac{1}{a}+\frac{r_0 k^2}{2} + \mathcal{O}(k^4).
\end{equation}
The zero-range theory corresponds to the case where the range of the potential is much smaller than the other typical length scales of the system, and thus the $\mathcal{O}(k^2)$ term in Eq.~(\ref{eq:delta}) can be neglected and we have a description in terms of only the scattering length. In situations where the range of the potential is small but non-negligible, we can include higher-order contributions by considering the effective range $r_0$. 

Equation~(\ref{eq:delta}) is often called shapeless or shape-independent approximation because higher-order terms depend on the shape of the two-body potential. The usefulness of this equation is that two different microscopic potentials, which can be of entirely distinct functional forms, yield the same low-energy phase shifts, provided that both have the same scattering 
length and effective range.

\begin{figure}[!htb]
    \centering
    \includegraphics[angle=-90,width=\linewidth]{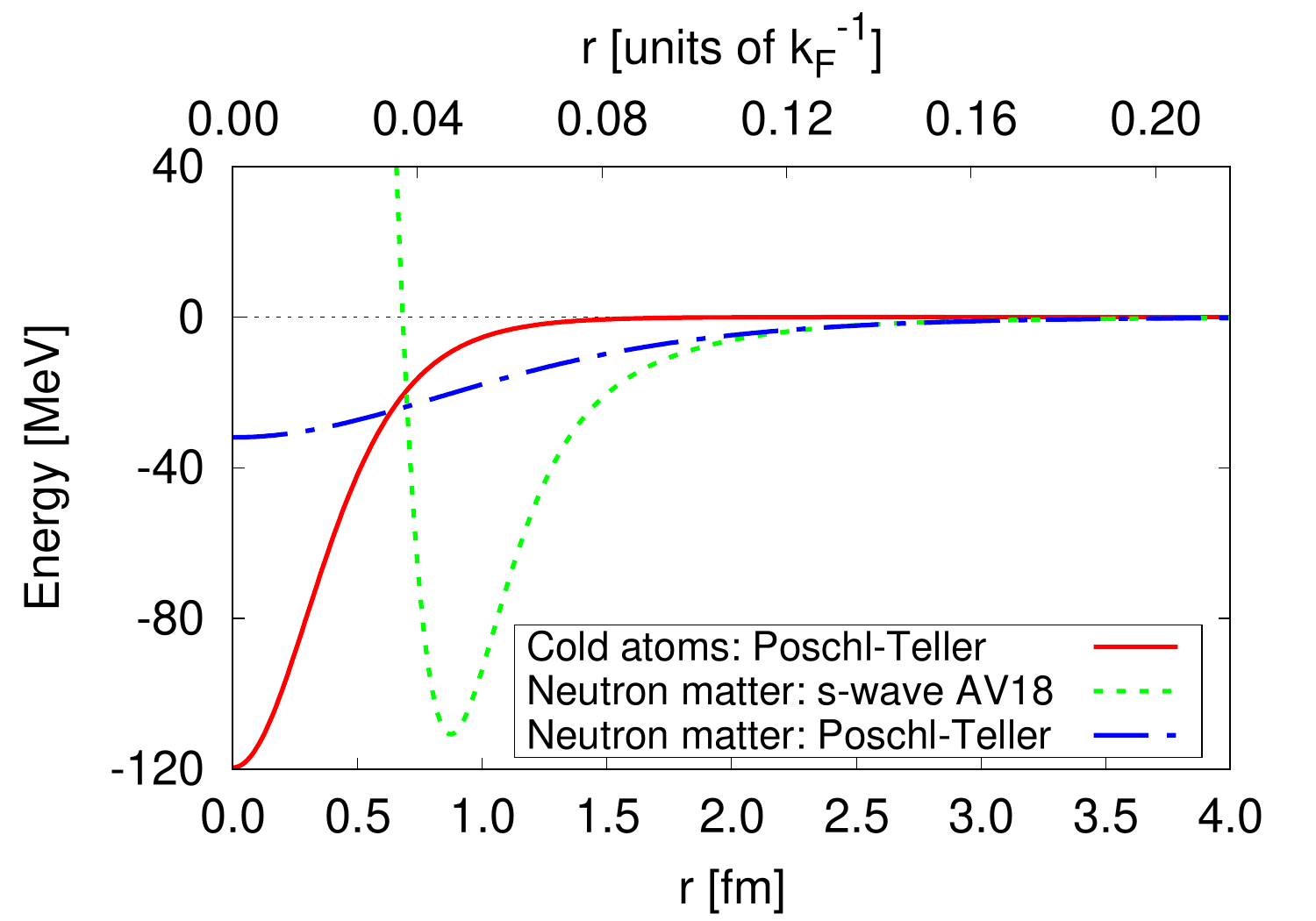}
    \caption{
Three different two-body potentials are depicted, corresponding to $-k_Fa=1$, where $k_F$ represents the Fermi wave number. The red curve represents the modified P\"oschl-Teller potential with a very small effective range ($k_F r_0=0.05$), the dashed green curve represents the $s$-wave component of AV18, and the dotted-dashed blue line represents the modified P\"oschl-Teller potential tuned to neutron-neutron low-energy parameters, $a^{nn}=-18.5$ fm and $r_0^{nn}=2.7$ fm. The top horizontal axis indicates the separation in units of $k_F^{-1}$, assuming the number density to be the same as the free Fermi gas ($n=k_F^3/(3\pi^2)$). Figure taken from Ref.~\cite{Madeira2019}.
}
    \label{fig:pot}
\end{figure}

In systems without a three-body scale, such as two-component Fermi gases, Eq.~(\ref{eq:delta}) has facilitated comparisons of results obtained with potentials of diverse shapes: square-well~\cite{Astrakharchik2004}, modified P\"oschl-Teller (mPT)~\cite{Carlson2003}, and the $s$-wave component of nuclear potentials~\cite{Gezerlis2008,Gezerlis2010,Madeira2019}. In Fig.~\ref{fig:pot}, we show different two-body potentials used in Ref.~\cite{Madeira2019} to investigate the universality of cold fermionic gases and low-density neutron matter. For the present discussion, we focus on the two potentials employed to model the neutron-neutron interactions: the $s$-wave component of AV18~\cite{Wiringa1995} and the modified P\"oschl-Teller potential tuned to the same scattering length and effective range. These two interactions differ considerably in shape: the former has a strong short-range repulsion and a weakly attractive tail, while the latter is purely attractive. However, since they reproduce the same scattering length and effective range, the low-energy properties investigated in Ref.~\cite{Madeira2019} using both potentials were in agreement, a consequence of Eq.~(\ref{eq:delta}).

\subsection{Loosely-bound dimers in physical systems}
\label{sec:dimers_physical}

The two-body $s$-wave scattering amplitude is given by~\cite{Newton2013}
\begin{equation}
    f(k)=\frac{1}{k\cot \delta_0-ik}\, .
\end{equation}
Its pole is related to the bound or virtual dimer energy
\begin{equation}
\label{eq:EB}
E_B=-\frac{\hbar^2}{2 m_r a_B^2},
\end{equation}
where $m_r$ is the reduced mass of the system and the binding length $a_B$ is related to the scattering length and effective range through~\cite{Jamieson1995,Janzen1995}
\begin{equation}
\label{eq:aB}
\frac{r_0}{a_B}=\frac{r_0}{a}+\frac{1}{2}\frac{r_0^2}{a_B^2}.
\end{equation}
We kept only the first two terms of the $k\cot\delta_0(k)$ effective range expansion, Eq.~(\ref{eq:delta}), in this expression.

Since Eq.~(\ref{eq:EB}) reduces to
\begin{equation}
\label{eq:Ezr}
E_{\rm zr}=-\frac{\hbar^2}{2 m_r a^2}
\end{equation}
in the zero-range limit (negligible effective range), we can compare both approximations to physical systems to better understand the impact of the effective range in their low-energy properties. In Table~\ref{tab:systems}, I summarized the scattering length and the binding length of two-body systems belonging to atomic and nuclear physics~\cite{Gutierrez1984,Budzanowski2010,Kievsky2021}. The closer these two quantities are, the less important is the range of the potential.

\begin{table}[!htb]
\caption{Scattering length and binding length of several two-body systems encompassing atomic and nuclear physics. The values are taken from Refs.~\cite{Gutierrez1984,Budzanowski2010,Kievsky2021}.
}
\begin{tabular}{l|c|c}
\label{tab:systems}
\textbf{System}                  	& $\bm{a}$   & $\bm{a_B}$  \\ \hline 
\multicolumn{3}{c}{\textbf{Atomic Physics (nm)}}  \\ \hline
$^4$He dimer               	& \phantom{$-$}9.04    &  \phantom{$-$}8.65   \\ \hline
$^3$He dimer               	& $-$0.70    & $-$1.13     \\ \hline
\multicolumn{3}{c}{\textbf{Nuclear Physics (fm)}} \\ \hline
deuteron [p-n (1$^+$)]           	& \phantom{$-$}5.42	& \phantom{$-$}4.32    \\ \hline
proton-neutron (0$^+$)     	& $-$23.74   			& $-$25.05    \\ \hline
neutron-neutron (0$^+$)    	& $-$18.90   			& $-$20.19  \\ \hline
p$\Lambda$ (singlet)      	& $-$2.43  				& $-$3.25    \\ \hline
p$\Lambda$ (triplet)      	& $-$1.56  				& $-$2.65   
\end{tabular}
\end{table}

\begin{figure}[!htb]
    \centering
    \includegraphics[height=4.0cm]{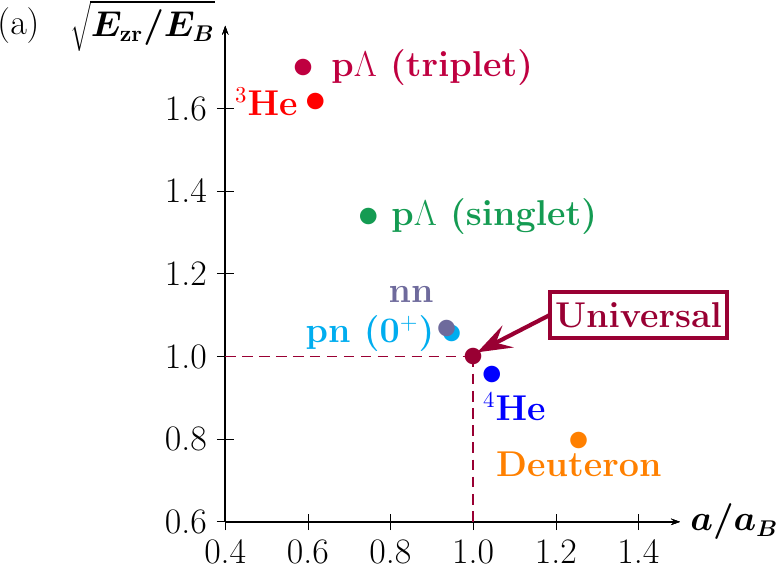}
    \hspace{0.5cm}
    \includegraphics[height=4.0cm]{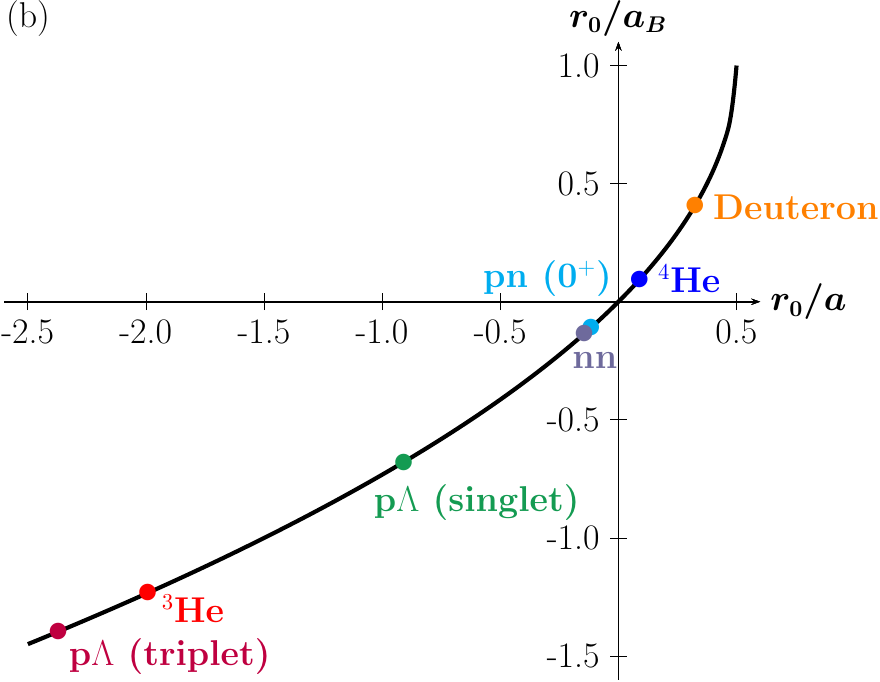}
    \caption{Physical systems of Table~\ref{tab:systems} and how they relate to the zero-range and shapeless universality regimes. Panel (a) considers only the scattering length, Eq.~(\ref{eq:Ezr}), and thus only a few systems are close to the universal zero-range limit of $a=a_B$. Panel (b) takes into account the effective range as the curve corresponds to Eq.~(\ref{eq:aB}). The shapeless universality is verified for these systems, as they are very close to the curve.}
    \label{fig:zero_finite}
\end{figure}

Figure~\ref{fig:zero_finite} is a graphical representation of Table~\ref{tab:systems} to illustrate how the selected physical systems relate to the zero-range and shapeless universality regimes. In panel (a), only the scattering length is taken into account, Eq.~(\ref{eq:Ezr}); hence the universal regime is when $a=a_B$. We can see that the $^4$He dimer, two neutrons, and the unbound state of a proton and a neutron (the $0^+$ channel) are close to this limit. However, the other systems are far from this regime, and the zero-range theory yields a crude description of their properties. In panel (b), I plot Eq.~(\ref{eq:aB}), which takes into account the effective range. We can see that all the considered physical systems are close to the curve, indicating that a description including the effective range for these systems is adequate.

\subsection{Illustration with a microscopic potential}
\label{sec:illustration_two_body}

When modeling the physical systems introduced in Sec.~\ref{sec:dimers_physical} with a microscopic potential, an explicit functional form must be chosen. Here, I present the zero-range and finite-range approximations of the dimer energy considering the modified P\"oschl-Teller potential, which has been successfully used to 
describe interactions in cold atom systems \cite{Carlson2003,Madeira2016,Madeira2017,Madeira2019,Madeira2021}. It can be written as
\begin{equation}
\label{eq:mPT}
V_{\rm mPT}(r)=- \frac{\hbar^2 \mu_{\rm PT}^2}{m_r} \frac{\lambda_{\rm PT}(\lambda_{\rm PT}-1)}{\cosh^2(\mu_{\rm PT} r)}.
\end{equation}
The potential is illustrated in Fig.~\ref{fig:pot} for two different sets of the parameters $\lambda_{\rm PT}$ and $\mu_{\rm PT}$, which are tuned to reproduce the desired scattering length and effective range.

This potential is a common choice since it is smooth, and there is an analytical expression that relates the parameters $\lambda_{\rm PT}$, $\mu_{\rm PT}$, and the scattering length~\cite{Madeira2019}. Moreover, in this work, I derived an analytical expression for the $s$-wave bound-state energies of this potential in three dimensions, 
\begin{equation}
    \label{eq:energy_levels_mpt}
    E_{\rm PT}=-\frac{\hbar^2\mu_{\rm PT}^2}{2m_r}(\lambda_{\rm PT}-2-2n)^2, \quad \left(n=0,1,2,... \text{ and } n\leqslant\frac{\lambda_{\rm PT}}{2}-1\right),
\end{equation}
where the derivation is presented in Appendix~\ref{sec:app_mpt}.

In Fig.~\ref{fig:dimer_cosh}, I compare the analytical result for the bound-state energies of the mPT dimer, obtained by choosing $n=0$ and $\lambda_{PT}>2$ in Eq.~(\ref{eq:energy_levels_mpt}), with the low-energy approximations. If the effective range is small compared to the scattering length ($r_0/a\ll 1$), then both approximations are in excellent agreement with the analytical solution. However, as the ratio $r_0/a$ increases, the zero-range approximation deviates more from the result than the finite-range one. Remarkably, the shapeless approximation yields results that differ by only a few per cent for $r_0/a \lesssim 0.1$.

\begin{figure}[!htb]
    \centering
    \includegraphics[angle=-90,width=0.7\linewidth]{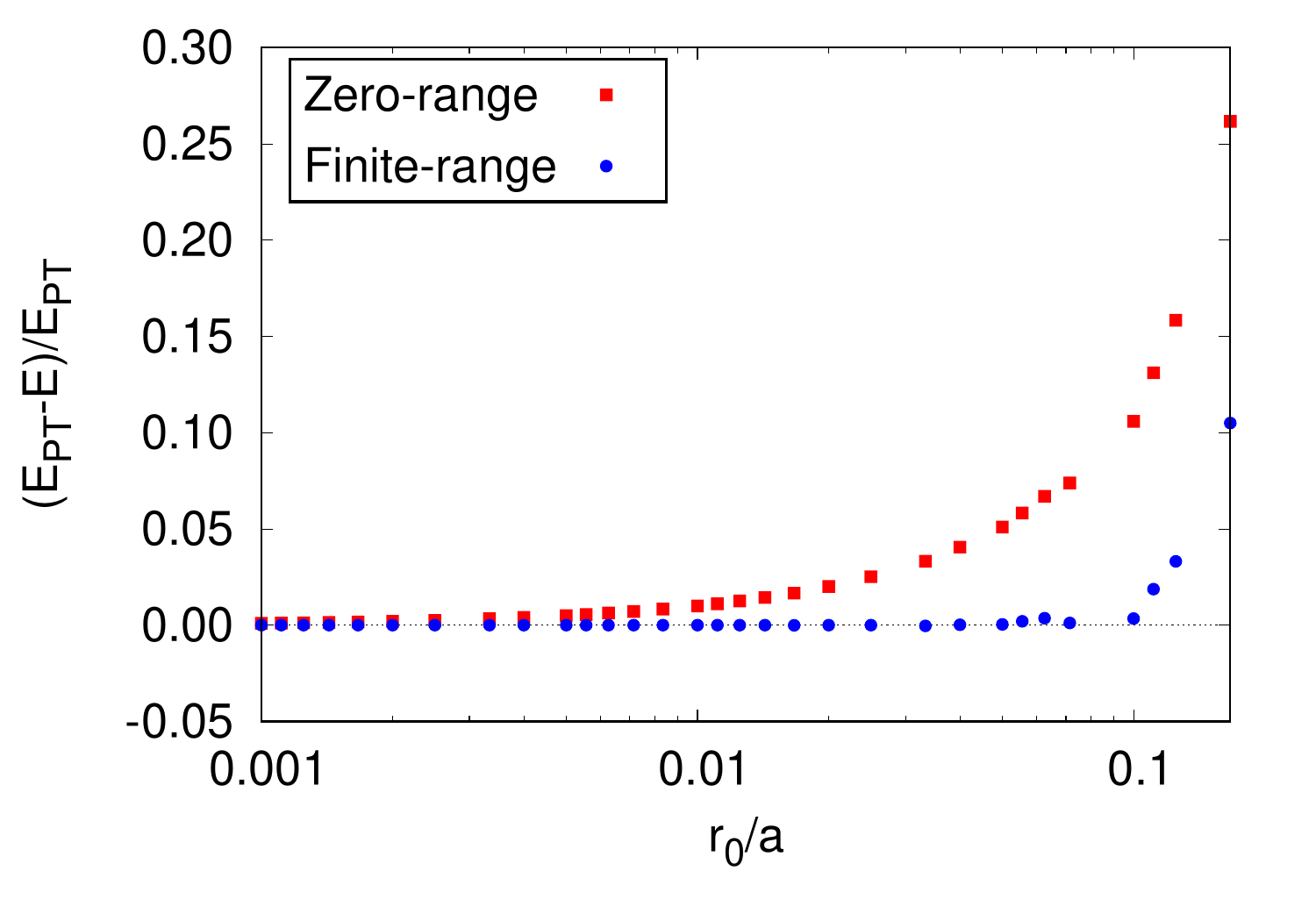}
    \caption{Comparison between the analytical dimer bound-state energy $E_{\rm PT}$ of the modified P\"oschl-Teller potential, Eq.~(\ref{eq:energy_levels_mpt}), and the zero-range and finite-range approximations, Eqs.~(\ref{eq:Ezr}) and (\ref{eq:EB}), respectively.}
    \label{fig:dimer_cosh}
\end{figure}

\section{Bosonic trimers}
\label{sec:three_body}

I discussed only two-body systems in Sec.~\ref{sec:two_body}. Although it is more challenging to consider range corrections systematically for $N>2$ systems, much progress has been made. Initially, linear range corrections were introduced to elucidate the Phillips line~\cite{EfimovTkachenko1985}, sparking numerous studies on range corrections across various systems~\cite{Efimov1991,
Hammer2001,Thogersen2008,Platter2009,Pricoupenko2010,Braaten2011,Castin2011,Werner2012,Wang2012,Tusnski2013,Ji2015,Souza2016}. These corrections have significant implications, especially with the ongoing experimental explorations in cold-atom physics~\cite{Wild2012,Makotyn2014,Chapurin2019,Xie2020,Zou2021}, highlighting the need for accurate expansion parameters near the unitary regime~\cite{Mestrom2019}. An approach that has yielded very interesting and promising results is a Gaussian parametrization of the universal region~\cite{Gattobigio2014,Kievsky2014,Kievsky2015,Rodriguez2016,Gattobigio2019,Gattobigio2019b,Kievsky2020,Recchia2022}.

In this section, I will focus on an approach to describe Efimov trimers through a universal energy scaling function that explicitly considers the effective range. Reference~\cite{Madeira2021} aimed to develop range corrections for a trimer of identical bosons near unitarity, using a universal scaling function to relate the trimer energies with different scattering lengths and effective ranges. This novel framework allows for systematic extensions to larger systems.

A three-body scale is needed to avoid the Thomas collapse in a three-boson system with a $s$-wave zero-range force, as the two-body scattering length alone does not suffice for determining the low-energy properties of the trimer. By incorporating corrections from the finite range through the effective range expansion and selecting a reference three-body energy at unitarity $E_3({1}/{a}=0,r_0,\nu)$ (where $\nu$ is a three-body scale), we can combine the scattering length and effective range to create two dimensionless quantities:
\begin{eqnarray}
\label{eq:x}
x = \frac{\hbar}{a\sqrt{-m E_3(0,r_0,\nu)}},\\
\label{eq:y}
y = \frac{r_0\sqrt{-m E_3(0,r_0,\nu)}}{\hbar}\,.
\end{eqnarray}
These definitions were constructed such that $x=0$ yields the unitary limit and $y=0$ the zero-range limit.

The energy scaling function $F(x,y)$ is defined as
\begin{eqnarray}
\label{eq:universal}
F(x,y)=\frac{E_3\left(1/a,r_0,\nu\right)}{E_3(0,r_0,\nu)}.
\end{eqnarray}
Extensive studies have been conducted on the zero-range limit of Eq.~(\ref{eq:universal}), presented in a different version that retains the same information~\cite{Braaten2006}. The calculation of binding energies with exceptional accuracy has been reported in several references~\cite{Braaten2006,Naidon2017,Mohr2006,Gattobigio2019b}. We aim to expand upon the zero-range limit by calculating trimer energies that account for finite effective ranges. While formulating an analytic expression for the scaling function in Eq.~(\ref{eq:universal}) poses difficulties, certain characteristics of it have been identified in the literature~\cite{Amorim1992,Castin2011,Werner2012}.

\begin{figure}[!htb]
\centering
\includegraphics[angle=-90,width=8.5cm]{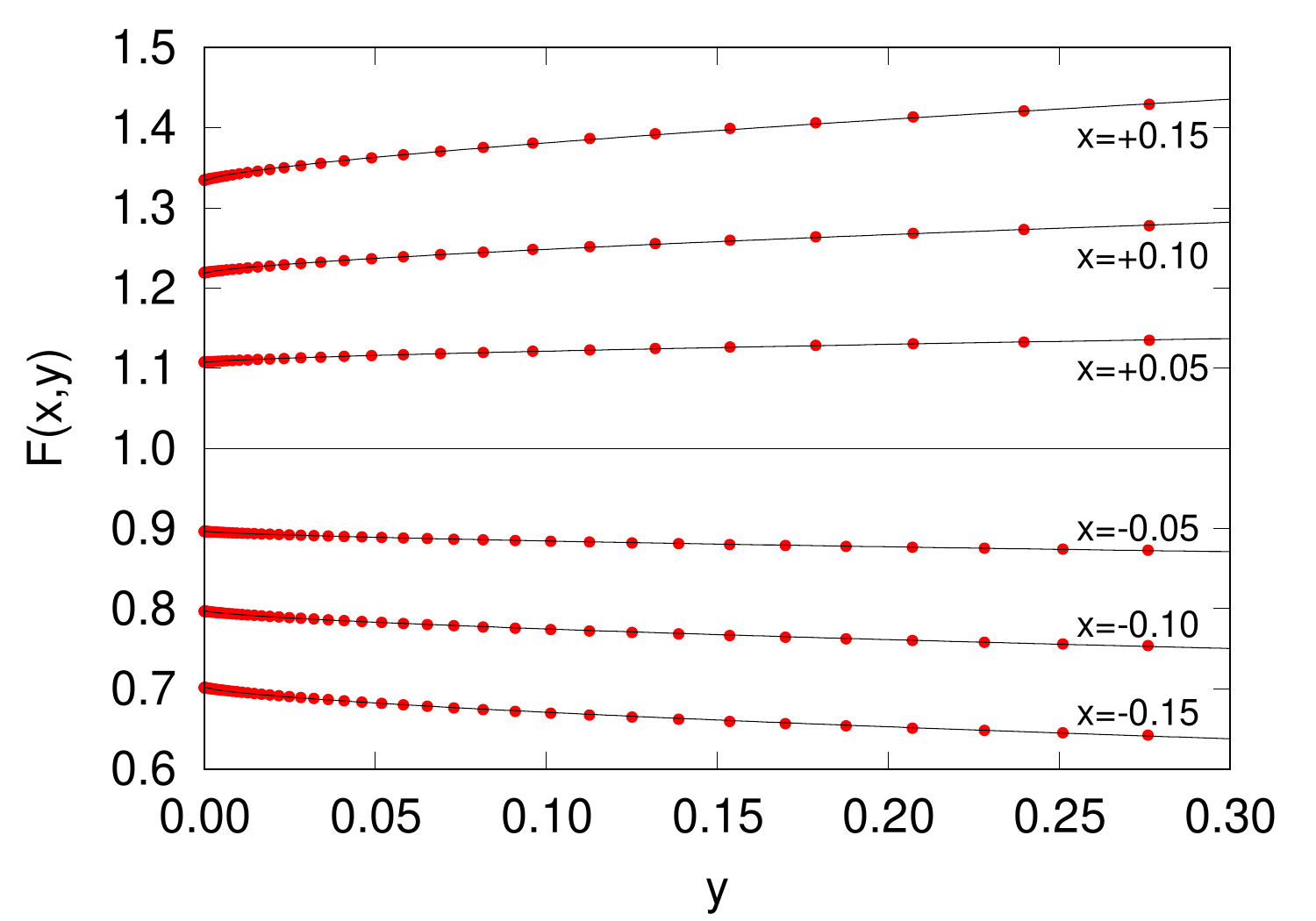}
\caption{
The function $F(x,y)$, representing the universal trimer energy scaling, is adjusted according to the functional form presented in Eq.~(\ref{eq:F}). The data points, shown as circles, are obtained from the STM equation with finite-range effects, whereas the lines depict the fitted model. Figure taken from Ref.~\cite{Madeira2021}.
}
\label{fig:fit}
\end{figure}

To obtain the scaling function, we utilized the solutions of the Skorniakov and Ter-Martirosian (STM) equation, incorporating the first-order effective range corrections as outlined in Ref.~\cite{Madeira2021}. The proposed expression of the scaling function is given by:
\begin{equation}
\label{eq:F}
F(x,y)=1+c_1 x+c_2 xy^\sigma+c_3 x^2+c_4 x^2y+c_5 x^2y^\sigma,
\end{equation}
representing a series expansion in terms of $x$ and $y$. The coefficients $c_i$ and the exponent $\sigma$ were obtained by fitting the STM data to this equation, with their specific values provided in Ref.~\cite{Madeira2021}. For comparison, the scaling function and the STM findings are illustrated together in Fig.~\ref{fig:fit}.

After determining the energy scaling function from the STM equation with effective range corrections, the next task was to compute it with microscopic two- and three-body interactions, which was done with quantum Monte Carlo (QMC) methods, as described in Ref.~\cite{Madeira2021}. The results indicated model dependence for large values of $y$, but they agreed with the scaling function for small values of $y$. The conditions that yield the universal behavior described by the scaling function were investigated, and the conclusion was that universal behavior was observed only if the size of the trimer was much larger than the range of the microscopic force, in agreement with what is expected from Efimov trimers.

It is important to note that the results using QMC techniques pertain exclusively to the ground state properties of the trimers. This offers a different strategy from the traditional method of investigating Efimov physics by examining excited states.

Although this work was restricted to the $N=3$ system, it would be interesting to construct analogous scaling functions for $N$-boson systems. An intriguing possibility is that considering the finite range effects of the interactions in these systems can restore what has been dismissed as nonuniversal behavior.

\section{Conclusion}
\label{sec:conclusion}

In this work, I motivated and presented a few examples of the investigation of finite-range effects in strong-interacting few-body systems. This included the discussion of concepts such as the zero-range and shapeless universalities in two-body systems. These are not just theoretical constructs, as they are relevant to studying a wide range of physical systems in atomic and nuclear physics.

The case of three identical bosons was also discussed, which leads to the remarkable Efimov effect. Although many works considered the effective range in this setting, I followed Ref.~\cite{Madeira2021}, where the developed formalism allows for systematically investigating finite-range effects in Efimov trimers.

Although many of the references provided in Sec.~\ref{sec:three_body} consider, in addition to $N=3$, small clusters ($N\lesssim 10$), there are not many studies in the literature where many-body strongly-interacting bosonic systems are investigated with low-energy universality or Efimov physics in mind. Recently, a quantum Monte Carlo study~\cite{Carlson2017} obtained the ground-state binding energies at unitarity for bosonic clusters with sizes much larger than the interaction range for up to $N=60$ and bulk properties. This opens up the possibility of studying finite range effects, which stem from few-body interactions, in many-body systems.

Besides being able to describe complex systems with just a few parameters, the importance of the low-energy universality is to connect fields that span several scales, from atomic to particle physics.

Recent progress in modeling strongly-interacting physical systems has shown that there are universal aspects shared by all systems close to unitarity. Still, an accurate quantitative description of a particular system has to include model-dependent features. One of the most critical roles of finite-range contributions is to increase the scope of the universal behavior so that the particularities of a specific physical system are minor corrections if compared to the strongly interacting universality. I hope this work motivates studies with this goal in mind.

\backmatter

\bmhead{Acknowledgements}
I thank the participants of the conference ``Critical stability of few-body quantum systems 2023'' for the fruitful discussions that inspired much of this manuscript. This work was supported by the S\~ao Paulo Research Foundation (FAPESP) under grant 2023/04451-9.

\begin{appendices}

\section{Analytical bound-state spectrum of the modified P\"oschl-Teller potential}\label{sec:app_mpt}

The modified P\"oschl-Teller potential, Eq.~(\ref{eq:mPT}), is one of the rare cases in quantum mechanics where we can obtain analytical solutions. It can be derived from supersymmetric quantum mechanics as the supersymmetric partner of the free particle potential~\cite{Sukumar1985,Diaz1999}. For our purposes, we are interested in seeing this potential as a smeared-out delta function, which is more convenient to implement in QMC schemes and other numerical approaches.

Almost all instances of analytical solutions involving the mPT deal with a single particle in one dimension, where $-\infty<x<+\infty$ \cite{Rosen1932,Landau1958,Alhassid1983,Sukumar1985,Diaz1999}. The appropriate boundary conditions, in this case, require that the wave function vanishes at $x=\pm\infty$. In this work, I employed the mPT potential as a two-body interaction in three dimensions. The differential equations of both cases are essentially the same if we employ center-of-mass coordinates in the 3D case and solve for the $s$-wave ($\ell=0$) reduced radial wave function $u(r)=rR(r)$. However, the boundary conditions differ; we must have $u(0)=u(r\to\infty)=0$.

I follow Refs.~\cite{Landau1958,Flugge1971} whenever possible. We want to solve the equation:
\begin{equation}
    u''+\left[\frac{\mu^2\lambda(\lambda-1)}{\cosh^2{(\mu r)}}+k^2\right]u=0,
\end{equation}
where $k^2=2m_rE/\hbar^2$, in the domain $0\leqslant r<\infty$, with the boundary conditions $u(0)=u(r\to\infty)=0$, $\mu>0$, and $\lambda>1$. First, I perform the substitution $y=\cosh^2(\mu r)$, which yields
\begin{equation}
    y(1-y)u''+\left(\frac{1}{2}-y\right)u'-\left[\frac{k^2}{4\mu^2}+\frac{\lambda(\lambda-1)}{4y}\right]u=0,
\end{equation}
with $1\leqslant y <\infty$. Next, I perform the transformation $u(y)=y^{\lambda/2}v(y)$,
\begin{equation}
    y(1-y)v''+\left[\left(\lambda+\frac{1}{2}\right)-(\lambda+1)y\right]v'-\frac{1}{4}\left(\lambda^2+\frac{k^2}{\mu^2}\right)v=0.
\end{equation}
I define:
\begin{eqnarray}
    \label{eq:complex_a}
    a&=&\frac{1}{2}\left(\lambda+\frac{ik}{\mu}\right),\\
    \label{eq:complex_b}
    b&=&\frac{1}{2}\left(\lambda-\frac{ik}{\mu}\right),\\
    c&=&\lambda+\frac{1}{2},
\end{eqnarray}
where I used the usual notation, and $a$ should not be confused with the scattering length. The differential equation can be written as
\begin{equation}
    y(1-y)v''+\left[c-(a+b+1)y\right]v'-abv=0,
\end{equation}
which is the hypergeometric differential equation (see Eq.~(15.5.1) of Ref.~\cite{Abramowitz1965}). If none of the numbers $c$, $c-a-b$, or $a-b$ is an integer, then two linearly independent solutions in the neighborhood of $y=1$ exist (see Eqs.~(15.5.5) and (15.5.6) of Ref.~\cite{Abramowitz1965}). The general solution is given by
\begin{equation}
    \label{eq:vy}
    v(y)=A\, F(a,b;1/2;1-y)+B\, (1-y)^{1/2}F(a+1/2,b+1/2;3/2,1-y),
\end{equation}
where $F$ is the hypergeometric function, sometimes denoted by $_2F_1$, and $A$ and $B$ are constants to be determined.

Expressing the solution [Eq.~(\ref{eq:vy})] in terms of the reduced radial wave function yields
\begin{flalign}
    &u(r)=A\, \cosh^\lambda{(\mu r)} F(a,b;1/2;-\sinh^2(\mu r))+\nonumber\\
    &B\, \cosh^\lambda{(\mu r)}(-\sinh^2(\mu r))^{1/2} F(a+1/2,b+1/2;3/2,-\sinh^2(\mu r)).
\end{flalign}
The $u(0)=0$ boundary condition implies that $A=0$. I choose $B=-i$ so that the solution is
\begin{equation}
    \label{eq:ur}
    u(r)=\cosh^\lambda{(\mu r)}\sinh(\mu r) F(a+1/2,b+1/2;3/2,-\sinh^2(\mu r)).
\end{equation}
Applying a linear transformation, see Eq.~(15.3.7) of Ref.~\cite{Abramowitz1965}, I can write:
\begin{flalign}
    \label{eq:linear_transf}
    &u(r)=\cosh^\lambda{(\mu r)}\sinh(\mu r)\times\nonumber\\
    &\left[
    \frac{\Gamma(3/2)\Gamma(b-a)}{\Gamma(b+1/2)\Gamma(1-a)}(\sinh(\mu r))^{-2a-1}F\left(a+1/2,a;1-b+a;\frac{-1}{\sinh^2(\mu r)}\right)\right.\nonumber\\
    &\left.
    +\frac{\Gamma(3/2)\Gamma(a-b)}{\Gamma(a+1/2)\Gamma(1-b)}(\sinh(\mu r))^{-2b-1}F\left(b+1/2,b;1-a+b;\frac{-1}{\sinh^2(\mu r)}\right)
    \right].
\end{flalign}

Since we are interested in bound states, it is convenient to take $k=i\kappa$ such that the bound-state energy is
\begin{equation}
    \label{eq:energy_kappa}
    E=\frac{\hbar^2k^2}{2m_r}=-\frac{\hbar^2\kappa^2}{2m_r}.
\end{equation}
The parameters $a$ and $b$, Eqs.~(\ref{eq:complex_a}) and (\ref{eq:complex_b}), become real,
\begin{eqnarray}
    a&=&\frac{1}{2}\left(\lambda-\frac{\kappa}{\mu}\right),\nonumber\\
    b&=&\frac{1}{2}\left(\lambda+\frac{\kappa}{\mu}\right).
\end{eqnarray}
To impose the boundary condition $u(r\to\infty)=0$, we need to investigate the asymptotic behavior of Eq.~(\ref{eq:linear_transf}). The first term inside the square brackets diverges as $\exp(+\kappa r)$, while the second behaves as $\exp(-\kappa r)$. Hence, a normalizable solution is only possible if the first term vanishes, given that $\kappa>0$.

\begin{figure}[!htb]
    \centering
    \includegraphics[width=0.7\textwidth,angle=-90]{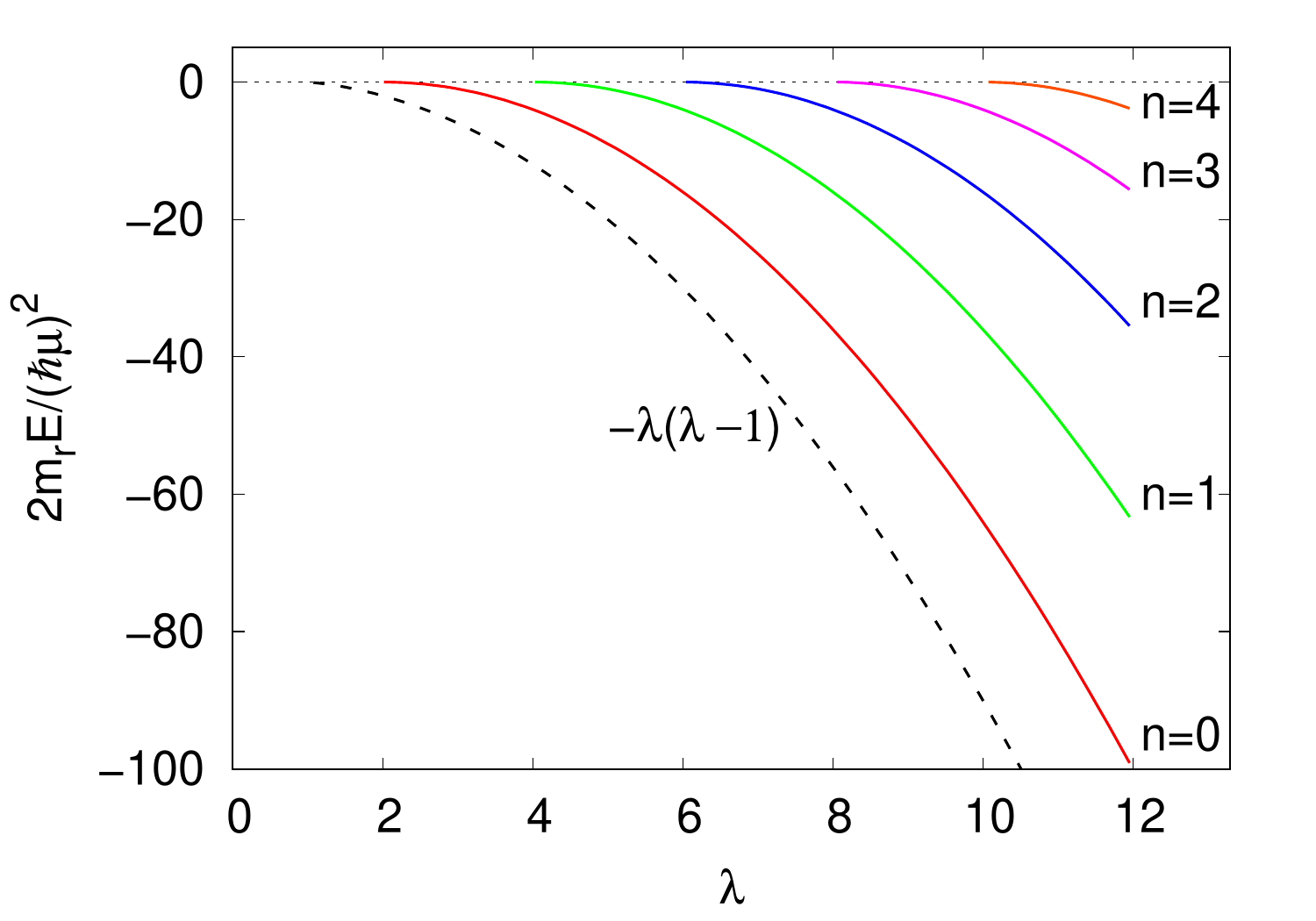}
    \caption{Energy levels of the three-dimensional two-body bound states of the modified P\"oschl-Teller potential, Eq.~(\ref{eq:bound_state_mpt}). The dashed line is associated with the depth of the potential.}
    \label{fig:mPT_bound_states}
\end{figure}

The $\Gamma$ functions in Eq.~(\ref{eq:linear_transf}) now have real arguments, since $a$ and $b$ are real. The $\Gamma$ function has poles at negative integers, which can be used to make the first term vanish. There are two $\Gamma$ functions in the denominator, which take as arguments $(b+1/2)$ and $(1-a)$. The first argument is never negative,
\begin{equation}
    b+\frac{1}{2}=\frac{\lambda}{2}+\frac{\kappa}{2\mu}+\frac{1}{2}>0.
\end{equation}
However, I can equate the second argument to negative integers,
\begin{equation}
    1-a=1-\frac{\lambda}{2}+\frac{\kappa}{2\mu}=-n, \quad (n=0,1,2,...).
\end{equation}
Solving for $\kappa$ and substituting into Eq.~(\ref{eq:energy_kappa}) yields
\begin{equation}
    \label{eq:bound_state_mpt}
    E=-\frac{\hbar^2\mu^2}{2m_r}(\lambda-2-2n)^2, \quad \left(n=0,1,2,... \text{ and } n\leqslant\frac{\lambda}{2}-1\right),
\end{equation}
which is the desired $s$-wave bound-state spectrum. It is consistent with the known property of the mPT potential that unitarity corresponds to $\lambda=2$ since, for this value, we only have a zero-energy state. In Fig.~\ref{fig:mPT_bound_states}, I illustrate the energy level dependence on the parameter $\lambda$.

\clearpage

\end{appendices}


\bibliography{references}

\end{document}